\documentclass[manuscript]{aastex}
\usepackage[T1]{fontenc}
\usepackage{amsmath}
\usepackage{amssymb}
\usepackage{tablefootnote}
\usepackage{color} 
\usepackage{xcolor}
\usepackage{soul}  

\begin{document}

\title{Electron Cooling in a Young Radio Supernova: SN 2012aw}
\author{Naveen Yadav\altaffilmark{1}$^\text{\dag}$, Alak Ray\altaffilmark{1}$^\text{\ddag}$,
Sayan Chakraborti\altaffilmark{2}, Christopher Stockdale\altaffilmark{3}, \\
Poonam Chandra\altaffilmark{4}, Randall Smith\altaffilmark{5}, Rupak Roy\altaffilmark{6},
Subhash Bose\altaffilmark{6}, Vikram Dwarkadas\altaffilmark{7},\\ Firoza Sutaria\altaffilmark{8} \&
David Pooley\altaffilmark{9}}

\affil{$^1$Tata Institute of Fundamental Research, Homi Bhabha Road, Mumbai $400005$, India}

\affil{$^2$Institute for Theory and Computation, Harvard Smithsonian Center for Astrophysics,\\
60 Garden Street, Cambridge, MA 02138, USA}

\affil{$^3$Marquette University, Milwaukee, WI 53233, USA}

\affil{$^4$National Center for Radio Astronomy-TIFR, Pune 411007, India}

\affil{$^5$Harvard-Smithsonian Center for Astrophysics, 60 Garden Street, Cambridge, MA 02138, USA}

\affil{$^6$Aryabhhata Research Institute of Observational Sciences, Nainital 263129, India}

\affil{$^7$ Department of Astronomy \& Astrophysics, University of Chicago, Chicago, IL 60637, USA}

\affil{$^8$Indian Institute of Astrophysics, Bangalore 560034, India}

\affil{$^9$ Department of Physics, Sam Houston State University, Huntsville, TX 77341, USA}

\email{$^\text{\dag}$nyadav@tifr.res.in,$^\text{\ddag}$akr@tifr.res.in}

\begin{abstract}
We present the radio observations and modeling of an optically bright Type II-P supernova (SN), SN 
2012aw which exploded in the nearby galaxy Messier 95 (M95) at a distance of $10\ \rm Mpc$. 
The spectral index values calculated using $C$, $X$ \& $K$ bands are smaller than the expected 
values for optically thin regime. During this time the optical bolometric light curve stays in the 
plateau phase. We interpret the low spectral index values to be a result of electron cooling. On 
the basis of comparison between Compton cooling timescale and Synchrotron cooling timescale we find 
that inverse Compton cooling process dominates over synchrotron cooling process. We therefore model 
the radio emission as synchrotron emission from a relativistic electron population with a high 
energy cutoff. The cutoff is determined by comparing the electron cooling time scale $t_{cool}$ and 
the acceleration time scale $\tilde t_{acc}$. We constrain the mass loss rate in the wind ($\dot 
M\sim 1.9\times10^{-6}\ \rm M_{\odot}yr^{-1}$) and the equipartition factor between relativistic 
electrons and the magnetic field  ($\tilde \alpha=\epsilon_e/\epsilon_B\sim 1.12\times10^2$) through 
our modeling of radio emission. Although the time of explosion is fairly well constrained by optical 
observations within about $2\ \rm days$, we explore the effect of varying the time of explosion to 
best fit the radio light curves. The best fit is obtained for the explosion date as 2012 March 15.3 
UT.
\end{abstract}

\section{Introduction}
Core-collapse SNe mark the death of massive stars ($\ \rm M_{*}/M_{\odot}\gtrsim8$). Study of 
electromagnetic emission from a supernova SN across various wavelengths provides us with important 
clues about the nature of the explosion as well as the progenitor star. Early time optical emission 
from a  SN is used to derive many important parameters of the explosion (e.g. total explosion 
energy, nickel mass etc.) whereas late time optical emission is a probe of the inner layers of 
ejecta. Though not all  SNe are detectable at radio wavelengths at a very young age, a small 
fraction of them have detectable radio emission even at a very young age. According to current 
understanding, this radio emission is non-thermal in origin \citep{1982ApJ...259..302C}. The fast 
moving SN ejecta drives a strong shock into the circumstellar medium (\emph{forward shock}). 
Electrons are accelerated to relativistic energies at this shock. These electrons gyrate around the 
post-shock magnetic field and radiate via synchrotron emission. This radiation is an important 
probe of the pre-explosion evolution of massive stars.

During their evolution, massive stars lose mass (by either continuous stellar winds or periods of 
rapid/episodic mass loss \citep{1984ApJ...287L..69D}) which forms the circumstellar medium in which 
the  SN shock evolves. The velocity of stellar winds is small (for Wolf-Rayet stars it can be $20$\% 
of the ejecta velocity) compared to that of the SN ejecta, and therefore in a short time the fast 
moving ejecta probes a long period of mass loss. Observationally determined mass loss rates can be 
used to constrain stellar evolution models. Young radio bright SN also offer an opportunity to 
study particle acceleration and magnetic field amplification at these shocks.

Type II-P SN are a class of core-collapse SN displaying an intermediate plateau phase in their 
bolometric light curve which extends from $60-100\ \rm days$.  They show a wide range of magnitude 
in plateau phase and expansion velocity \citep{2003ApJ...582..905H}. Their progenitor stars 
have an extended Hydrogen envelope prior to collapse \citep{2009ARA&A..47...63S}. Therefore 
they are at the extremity of a range of stars retaining different Hydrogen envelope masses at the 
time of explosion. As a result of the SN explosion the Hydrogen envelope is ejected at high 
velocity. The plateau phase is powered by a Hydrogen recombination wave traveling inwards as this 
ejecta cools due to expansion and radiation losses. The photosphere demarcates this expanding 
Hydrogen envelope into an inner region of high opacity and an outer region of low opacity. The 
plateau phase has been modeled numerically \citep{1983Ap&SS..89...89L,2011ApJ...729...61B}, 
semi-analytically \citep{1977ApJS...33..515F} and analytically 
\citep{1980ApJ...237..541A,1991SvAL...17..210C,1993ApJ...414..712P}. The extended duration of 
plateau phase makes these  SN more easily detectable even in low cadence surveys. The long duration 
of the plateau phase may have consequences for the non-thermal radiation processes. The high 
radiation density  of optical (UBVRI) photons during the plateau phase may cause effective cooling 
of relativistic electron population at the forward shock \citep{2006ApJ...641.1029C}.

X-ray emission from a young Type II-P  SN can be thermal or non-thermal in origin 
\citep{1982ApJ...259..302C}. The thermal component can originate as a result of free-free 
emission in the post-shock region or at the \emph{reverse shock} (a shock which is driven in to the 
expanding SN ejecta), whereas the non-thermal emission can be due to inverse Compton scattering of 
low energy photons  by relativistic electrons at the forward shock. Therefore in case a  SN is 
bright and detectable in X-rays at early times, much more information is available for understanding 
the dynamics of the forward shock, the reverse shock, the density profile of the ejecta and the 
circumstellar medium. In case of SN 2004dj, \citet{2012ApJ...761..100C} have estimated various 
important parameters relevant to blast wave dynamics and particle acceleration using $4$ epochs of 
\emph{Chandra} observations. In the case of SN 2011ja, \citet{2013arXiv1302.7067C} have reported 
that the X-ray flux from this  SN on second observation epoch was higher compared to the X-ray flux 
on first epoch by a factor of $4.2$. They have argued that it can be explained by an enhancement in 
the density of the circumstellar medium probed by the shock at later time and have suggested that a 
fraction of Type II-P explosions may take place inside bubbles blown by hot winds or variable 
circumstellar medium created by non-steady winds. Therefore following the temporal evolution of 
young Type II-P explosion in radio and X-rays band will provide us with crucial information about 
the explosion and their surrounding medium created during the late evolution of their progenitor 
stars.

SN 2012aw is a bright Type II-P  SN which exploded in the galaxy M95 (d $\sim 10\ \rm Mpc$). 
Spectra taken $4-5\ \rm days$ after discovery showed it to be a Type II-P explosion 
\citep{2012CBET.3054....1F}. \citet{2012ApJ...759L..13F} identified a candidate progenitor star in 
archival HST images. \citet{2012ApJ...759L..13F} have inferred a progenitor mass in range $14-26\ 
\rm M_\odot$, whereas \citet{2012ApJ...756..131V} inferred a progenitor mass in range $17-18 \ \rm 
M_\odot$. The progenitor seems to be a faint red super-giant and is the most massive Type II-P 
progenitor discovered till date. Both works noted that the star had a significantly higher 
extinction prior to its explosion as a  SN and interpret it as a signature of dust destruction by
explosion. \citet{2012ApJ...759L..13F} note that the progenitor's luminosity is not very well 
constrained because of uncertainty in the extinction which will further affect the estimates on 
progenitor's mass. \citet{2012ApJ...756..131V} claim evidence for dust-destruction by explosion as 
the current extinction to the  SN is very low. This may have interesting consequences for the 
progenitors of Type II-P SN. SN 2012aw has been extensively studied through optical and UV 
photometry. \citet{2013MNRAS.tmp.1583B} have found that SN 2012aw has remarkable similarities with 
SNe 1999em, 1999gi and 2004et.  \citet{2013MNRAS.tmp.1583B} have reported nebular spectroscopy of  
SN at age of $270\ \rm days$ and on the basis of lines profile shapes claimed that there are no 
signs of fresh dust formation. \citet{2012ATel.3995....1I} reported the detection of an X-ray 
point-source consistent with the optical position of the SN 2012aw, with a $3.8\ \sigma$ 
significance. We triggered the $K$ band radio observation of SN 2012aw under our Joint Chandra-EVLA 
proposal (Proposal No. 13500809) to observe bright and nearby Type II-P events. After the initial 
detection \citep{2012ATel.4010....1Y} the \emph{JVLA} radio follow up was carried out through 
\emph{Jansky VLA} Director's Discretionary Time. We have observed the object at radio wavelengths 
using JVLA\footnote{The National Radio Astronomy Observatory is a facility of the National Science 
Foundation operated under cooperative agreement by Associated Universities, Inc.} and 
GMRT\footnote{We thank the staff of the GMRT that made these observations possible. GMRT is run by 
the National Center for Radio Astrophysics of the Tata Institute of Fundamental Research.}, 
targeting it at {\it L} ($1.4\ \rm  GHz$), {\it S} ($3.0\ \rm GHz$), {\it C} ($5.0\ \rm  GHz$), 
{\it X} ($8.5\ \rm  GHz$), {\it K} ($21.0\ \rm  GHz$) \& {\it Ka} ($32.0\ \rm  GHz$) bands at 
multiple epochs. In this work we present the analysis and modeling of radio observations of this  
SN. We model the radio observations using the circumstellar interaction model.  We show that there 
is a signature of electron cooling in the spectral evolution of the  SN especially  at high 
frequencies. In our model we modify the electron population by taking inverse Compton cooling 
process in to consideration. We constrain the parameters relevant to progenitor (mass loss rate, 
$\dot M/v_w$, where $v_w$ is the wind velocity) and properties of shock acceleration (equipartition 
factor, $\epsilon_e/\epsilon_B$, where $\epsilon_e$ \& $\epsilon_B$ are the fraction of energy in 
relativistic electrons and post-shock magnetic field).

\section{Radio Observations \& Reduction}
SN 2012aw was first detected in radio JVLA-$K$ band ($21\ \rm GHz$) at $\sim10$ days 
by \citet{2012ATel.4010....1Y} \& \citet{2012ATel.4012....1S}. We conducted  the follow up radio 
observations of 2012aw at various epochs extending up to $184$ days after the explosion using 
\emph{Karl G. Jansky Very Large Array (JVLA)} and \emph{Giant Meterwave Radio Telescope (GMRT)}.
These observations have been reduced using Astronomical Image Processing Software ({\it AIPS}) 
standard techniques. Group delay and phase rates calibration were determined using \emph{AIPS} task 
\emph{FRING}. Noisy data was flagged and the interferometric visibilities have been calibrated 
using 3C286. Bandpass calibration was done using \emph{BPASS} based on the strong flux calibrators.
The single source data has been extracted using \emph{AIPS} task \emph{SPLIT} after 
final calibration. The single source data was imaged using \emph{IMAGR}. The images were corrected 
for residual calibration errors using self-calibration of visibility phases 
\citep{1989ASPC....6..185C}. The source fluxes were extracted by fitting Gaussian using task 
\emph{JMFIT} assuming point sources. The errors reported on the flux are obtained by using the 
image statistics from the region surrounding the source.

In the case of SN 2012aw explosion date is strongly constrained to within $\pm1.6 \ \rm days$ based 
on a non-detection (limiting magnitude of $\rm R \gtrsim 20.7$) on $\text{Mar}\ 15.27$ reported by 
\citet{2012ATel.3996....1P} and the first detection on $\text{Mar}\ 16.9$ reported by 
\citet{2012CBET.3054....1F}. We have used the explosion date as $\text{Mar}\ 16.1$ in this work. We 
have explored the effect of varying the explosion date within the $1.6 \ \rm days$ time range.
The radio observations are presented in Table \ref{Table-1: Radio Obs}. 

\section{Modeling the Radio Observations}
The interaction of fast moving ejecta with the circumstellar medium drives a strong shock which 
moves ahead of the ejecta into the circumstellar medium and is called the `forward shock'. Electrons 
are accelerated to relativistic energies at this shock via Fermi first order process. These 
electrons radiate via synchrotron mechanism in the post-shock magnetic field. The electron spectrum 
is described as
\begin{equation}
 N(E)= N_0E^{-\gamma}
\end{equation}
where $N_0$ is the normalization constant and $\gamma$ is electron index. The radio emission from 
young SNe is generally modeled as synchrotron emission by this electron population affected by a 
variety of absorption processes. The absorption can be modeled as a combination of synchrotron self 
absorption (SSA, the electrons which are responsible for synchrotron emission also absorb the 
synchrotron photons) \& free-free absorption (FFA, the thermal electrons in the post-shock medium 
absorb the synchrotron photons). We use Chevalier model-I (Table 1. \citet{1996ASPC...93..125C}) to 
study this emission. In this model the radius of the forward shock increases as, $R\propto t^m$ and 
energy densities in relativistic electrons and magnetic field are proportional to the thermal energy 
density which leads to $u_e,u_B\propto t^{-2}$, where $u_e$ is the energy density in the 
relativistic electrons and and $u_B$ is the energy density in the post-shock magnetic field 
respectively. Another important assumption inherent to the model is that the electron index $\gamma$ 
remains constant during the evolution. Electron index can be obtained by fitting a power law to the 
optically thin component. The equation for the radio flux evolution in such a case is given in 
\citet{1998ApJ...499..810C} for the case of a  SN blast wave expanding into a circumstellar medium 
set up by a uniform wind ($\rho_w\propto r^{-2}$).
If we try to model the radio emission from SN 2012aw by a simple \emph{SSA+FFA} model, the best fit 
gives $\chi_\nu^2\sim 7.2$, but results in a value of $m$ greater than $1$ (\emph{SSA} 
model: $m=1.1\pm0.02$ ; \emph{SSA+FFA} model: $m=1.08\pm0.02$), implying an accelerated 
blast wave, 
which is unlikely as the blast wave decelerates due to its interaction with the circumstellar 
matter. The difference between model and data at early time is relatively large compared to that at 
late times.

\begin{figure}
\begin{center}
 \includegraphics[width=0.8\textwidth]{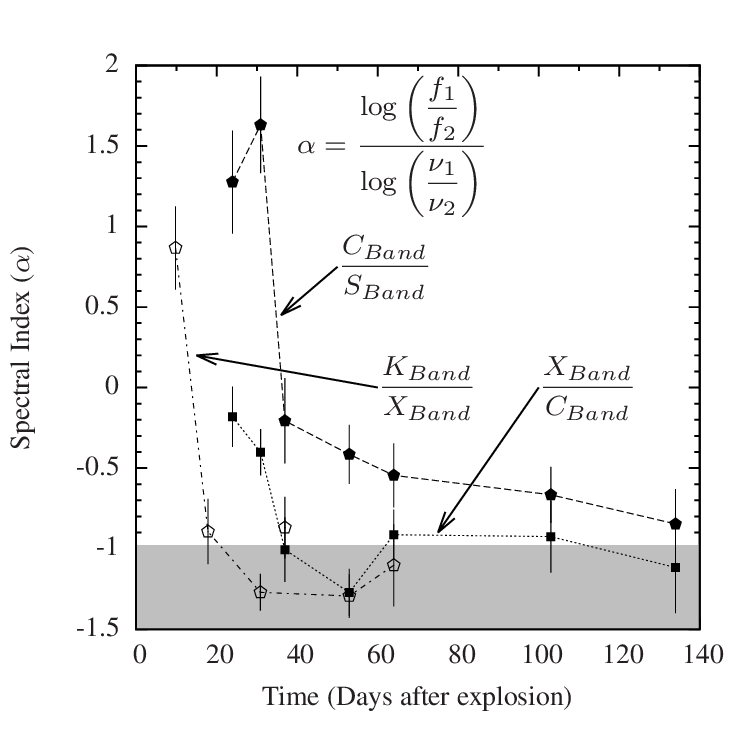}
\caption{Spectral index curves made using the $S$, $C$ , $X$ \& $K$ band data.
In case of the $X_{Band}/C_{Band}$ \& $K_{Band}/X_{Band}$ the spectral index values are smaller 
than $-1$ for an extended period of time between $20$ - $60\ \rm Days$. The $C_{Band}/S_{Band}$ 
index approaches the optically thin value slowly as the supernova ages. The spectral index 
calculated from higher frequency bands show observable deviations due to cooling of the 
relativistic 
electrons.}
\label{fig:spectral_index}
\end{center}
\end{figure}

In order to explore it further we  make a study of spectral index evolution using our radio data
as shown in Figure \ref{fig:spectral_index}. In case of a source that can be described by a simple
\emph{SSA+FFA} model without cooling the radio spectral index approaches the value, $-(\gamma -1)/2$
, as the source enters the optically thin regime. The spectral index curves labeled as 
`$X_{Band}/C_{Band}$' and `$K_{Band}/X_{Band}$' have values lower than $-1$ for an extended period 
of time during which the supernova has a plateau in its optical bolometric light curve, whereas 
`$C_{Band}/S_{Band}$' spectral index values slowly approach the optically thin regime value.
This is because due to electron cooling the flux in higher frequency bands is diminished more in 
comparison to lower frequency bands and this leads to a dip in the spectral index. 
The simplistic model proposed here may not fully account for the dip in the spectral index 
curves, -indicating that one may need to go beyond simple model described here to accommodate early 
time high frequency observations. A more realistic model will include the effect of variation in 
electron index and mass loss in to consideration as has been done in the case of SN 1993J by 
\citet{1998ApJ...509..861F}.

Electron cooling can be due to Coulomb, synchrotron or inverse Compton mechanisms or adiabatic 
expansion. Cooling has been discussed in the case of Type II-P  SNe by \citet{2006ApJ...641.1029C} 
and \citet{2004ApJ...605..823B} have discussed its importance in case of SN 2002ap, a type Ic event. 
To determine the dominant cooling mechanism, we need to compare the cooling timescales for various 
mechanisms.

\section{Cooling Timescales}
The rate at which an electron of energy $E$ loses energy by adiabatic expansion, inverse Compton 
scattering and synchrotron emission is \footnote{Details of of energy loss formula is 
given in \citet{1986rpa..book.....R} and for the case of a supernova (SN 1993J) 
by \citet{1998ApJ...509..861F}.}
\begin{eqnarray}
 &&\left(\dfrac{dE}{dt}\right)_{AD}\approx \dfrac{2}{3}Et\\
 &&\left(\dfrac{dE}{dt}\right)_{IC}\propto u_{rad}E^2\\
 &&\left(\dfrac{dE}{dt}\right)_{SC}\propto B^2E^2
\end{eqnarray}
respectively, where $u_{rad}$ is energy density of the radiation field and $B$ is the magnetic 
field. The characteristic energy loss timescale $t$ can be written as $E/\dot E$. The adiabatic 
cooling timescale $t_{ad}\propto t$ \citep{1982ApJ...259..302C}. The cooling timescales for inverse 
Compton and synchrotron therefore can be written using formulas for energy loss from 
\citet{1970ranp.book.....P} as
\begin{eqnarray}
t_{IC}&=&\dfrac{1}{3.97\times 10^{-2}u_{rad}E}\\
t_{SC}&=&\dfrac{1}{5.95\times 10^{-2} u_B E}
\end{eqnarray}
where $u_{rad}$ in our case is energy density of photons at  supernova radiosphere and $u_B$ is
energy density of the post-shock magnetic field. In the following subsections we will compare the 
cooling timescale for inverse Compton and synchrotron loses and determine the dominant cooling 
mechanism. In order to compare the cooling timescales we first need to estimate the post-shock 
magnetic field and the radiation density at forward shock.
\subsection{Post-Shock Magnetic Field}
To get the synchrotron cooling timescale we need an estimate of magnetic field. In 
the CSM interaction models for radio supernova, the post-shock magnetic field is assumed to scale 
with time according to a power law. In Chevalier model-I the magnetic field evolves as $t^{-1}$.
This is because magnetic energy density is proportional to thermal energy density, which for a 
constant parameter wind medium goes as $t^{-2}$, therefore $B\propto t^{-1}$.  
If we know the magnetic field at epoch $t_0$ it can simply be scaled to get the field at any other 
epoch using
\begin{equation}
B(t)=B_0\left(\dfrac{t}{t_0}\right)^a  \label{eqn-magnetic_field}
\end{equation}
We use the value $a=-1$ in our calculations in accordance with Chevalier model-I. To
get an estimate of magnetic field we can either use a late time radio spectrum or a low frequency 
radio lightcurve which are relatively free from the electron cooling effects. We consider the $3\ 
\rm GHz$ lightcurve for this part of calculation. In order to have minimum free parameters we need 
to check whether FFA is important to model the $3 \ \rm GHz$ radio data available to us ($t>23 \ 
\rm days$).

\begin{figure}
\begin{center}
 \includegraphics[width=0.8\textwidth]{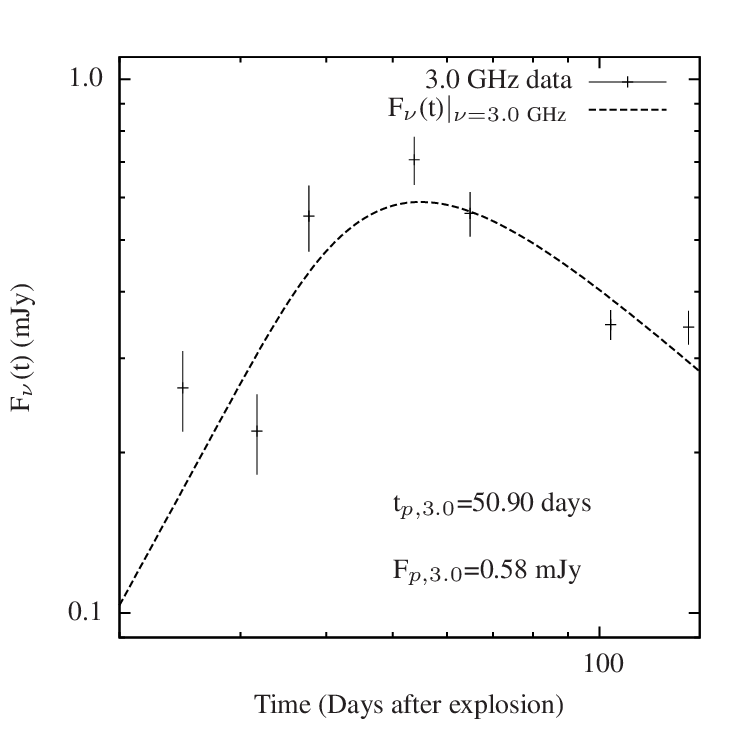}
\caption{An SSA model fit to the $3.0\ \rm GHz$ radio lightcurve using electron index $\gamma=3.1$.
The radio lightcurve in $3.0 \ \rm GHz$ band peaks at $50.90\ \rm days$ with the peak flux density
of $0.58\ \rm mJy$ and $m=0.97$. The $\chi^2_{\nu}\sim 5.4$ for the fit.} \label{fig:3GHz_lc}
\end{center}
\end{figure}

\begin{figure}
\begin{center}
 \includegraphics[width=0.65\textwidth]{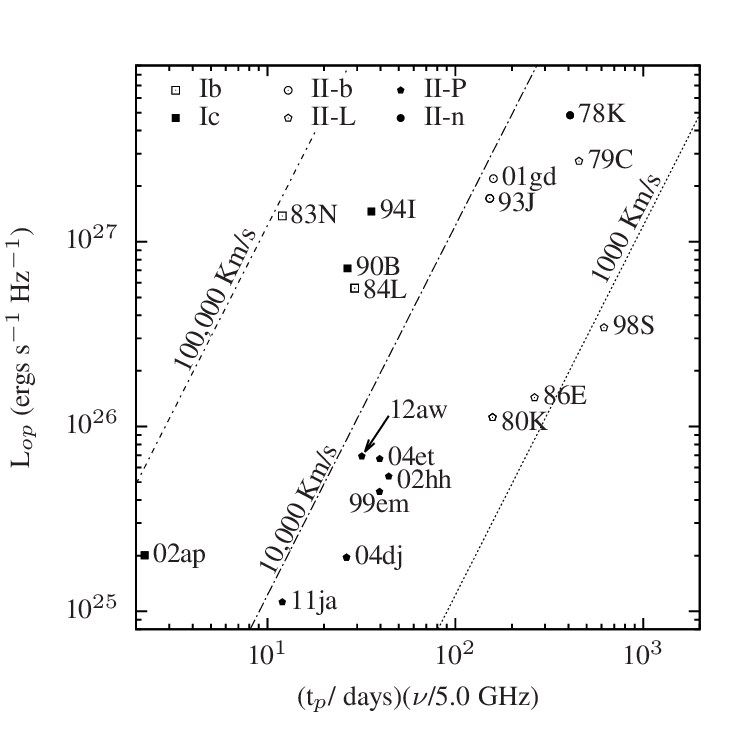}
 \includegraphics[width=0.65\textwidth]{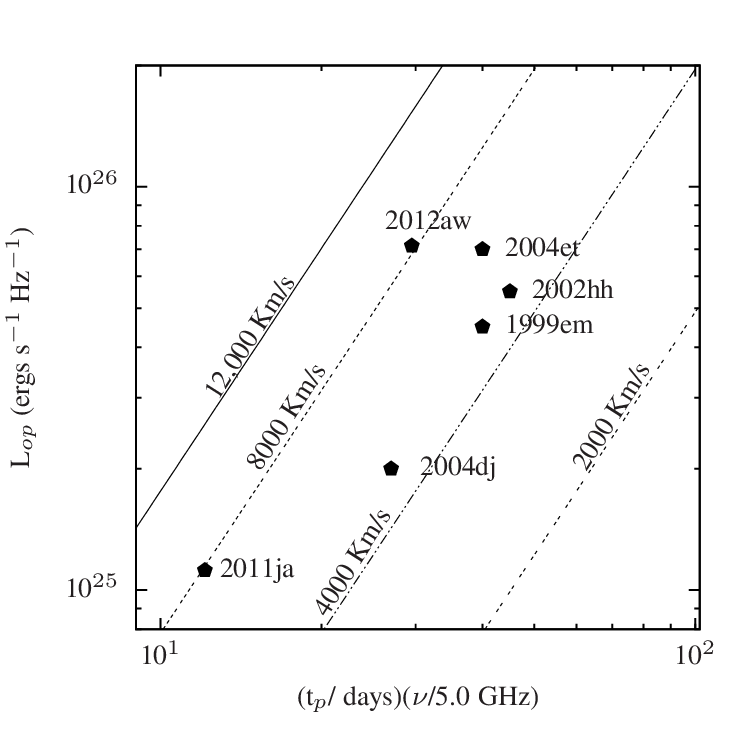}
 \caption{{\bf Upper Panel}: Type II-P supernova shown in the $L_{op}-\nu_pt_p$ plot(after 
\citet{1998ApJ...499..810C}) relative to the other core collapse SNe. {\bf Lower Panel}: SN 2012aw 
placed on a magnified $L_{op}-\nu_pt_p$ plot along with other radio Type II-P SNe. The constant 
velocity lines have been calculated using  electron index $\gamma=3.0$.} \label{fig:chev-diag}
\end{center}
\end{figure}

To get an estimate of the FFA we use $\dot M_{-5}/v_{w1}$ determined from epoch of X-ray detection 
(time at which the optical depth to X-rays becomes unity) where $\dot M_{-5}$ is mass loss rate in 
units of $10^{-5} \ \rm M_\odot\ yr^{-1}$, $v_{w1}$ is wind velocity in units of $10\ \rm km\ 
s^{-1}$. This object was first detected in X-ray ($0.2-10\ \rm KeV$ band) by 
\citet{2012ATel.3995....1I} approximately $4 \ \rm days$ after the explosion. This has been used to
get an upper limit on the quantity $\dot M_{-5}/v_{w1}$ which characterizes mass loss by a uniform 
wind. Using  Equation 2.17 from \citet{1994ApJ...420..268C}, we get
\begin{equation}
 \dfrac{\dot M_{-5}}{v_{w1}} =\dfrac{8.64\times 10^4 t_{X}v_{s4}E^{8/3}_{KeV}}{C_5} 
 \label{eqn:mass_loss}
\end{equation}
where $v_{s4}$ is outer (forward) shock velocity in units of $10^{4}\ \rm km\ s^{-1}$ and $t_{X}$ 
is the time at which the medium becomes optically thin to X-rays of energy $E_{KeV}$ and $C_5$ is a 
constant. Using $t_{X}=4 \ \rm days$, $v_{s4}\sim 1.0$, $E_{KeV}=1.0$ (\citet{2012ApJ...759...20K} 
find that there are no clear detections at low energies ($0.2-0.5\ \rm keV$) and only marginal 
detections at high energies ($2-10\ \rm keV$) $-$the observed counts are completely dominated by 
the $0.5-2\ \rm keV$ band) and substituting the value $C_5=2.6\times 10^6$ into Equation 
\ref{eqn:mass_loss}
\begin{equation}
  \dfrac{\dot M_{-5}}{v_{w1}}<0.13
\end{equation}
This is used to get an upper limit on the time for which free-free absorption dominates at any 
radio frequency. Using Equation 4 from \citet{2006ApJ...641.1029C}
\begin{equation}
 t_{ff}\approx 6\left(\dfrac{\dot M_{-6}}{v_{w1}}\right)^{2/3}T_{cs5}^{-1/2}v_{s4}^{-1}
 \left(\dfrac{\nu}{8.46\ \rm  GHz}\right)^{-2/3}
\end{equation}
where $t_{ff}$ is the time when the free free opacity becomes low enough so that the medium becomes
transparent to radio waves and $T_{cs5}$ is the circumstellar temperature in units of $10^5\ \rm K$.
This gives $t_{ff}\le 16.0\ \rm days$ at $3.0 \ \rm  GHz$ and $t_{ff}\le 11.0\ \rm days$ at $5.0 \ 
\rm  GHz$ for $T_{cs5}=1.0$. This shows that the $3\ \rm GHz$ radio lightcurve is not dominated by 
FFA in its optically thick phase (because our $3\ \rm GHz$ radio observations start from $23\ \rm 
days$ after explosion whereas $t_{ff}<16\ \rm days$). The $3.0 \ \rm GHz$ light curve can thus be 
fitted by a pure SSA model (Equation 4, \citet{1998ApJ...499..810C}) as shown in Figure 
\ref{fig:3GHz_lc}. The fitted value of $m$ is found to be $\sim0.97$ for the explosion date: 2012 
Mar 15.3 UT. A change in the assumed explosion date leads to differing values of the best fit $m$. 
The peak radio flux and the time to peak can be used to estimate the value of radius and magnetic 
field strength. Using the Equation $11$ \& $12$ from \citet{1998ApJ...499..810C} gives $B_0\sim0.48\ 
\rm Gauss$ and $R_0\sim 3.9\times 10^{15}\ \rm cm$ at age of $\sim 50.9 \ \rm days$ assuming 
equipartition. The magnetic field and assuming a different value of equipartition factor ($\tilde 
\alpha=\epsilon_e/\epsilon_B$) can be written as
\begin{eqnarray}
 B_0(\tilde \alpha)&=&0.46 \tilde \alpha^{-4/(2\gamma+13)} \ \rm Gauss\\
 R_0(\tilde \alpha)&=&4.9\times 10^{15} \tilde \alpha^{-1/(2\gamma+13)}\ \rm cm
\end{eqnarray}
We can now put the object on a $L_{op}-\nu_pt_p$ plot as shown in Figure \ref{fig:chev-diag} to 
compare it with the known Type II-P SNe. The object has a higher expansion velocity among known 
radio bright. The $L_{op}$ \& $\nu_pt_p$ values for SN 1999em, 2002hh, 2004et \& 2004dj have been 
taken from \citet{2006ApJ...641.1029C} and the values for SN 2011ja have been taken from 
\citet{2013arXiv1302.7067C}. The plot has been generated for electron index $\gamma=3.0$. SN 2012aw 
falls on a constant velocity line at around $8.0\times 10^3\ \rm km\ s^{-1}$ which is typical value 
of the blast wave speed. It also shows that the object is not much affected by FFA which is 
consistent with the low mass loss rate suggested by X-ray detection. The seemingly slow objects 
between $4.0\times 10^3\ \rm km\ s^{-1}$ and $8.0\times 10^3\ \rm km\ s^{-1}$ lines are dominated by 
FFA at early times or are affected by cooling at early times.

\subsection{Radiation Density and Bolometric Light Curve}
\begin{figure}
 \begin{center}
  \includegraphics[width=0.8\textwidth]{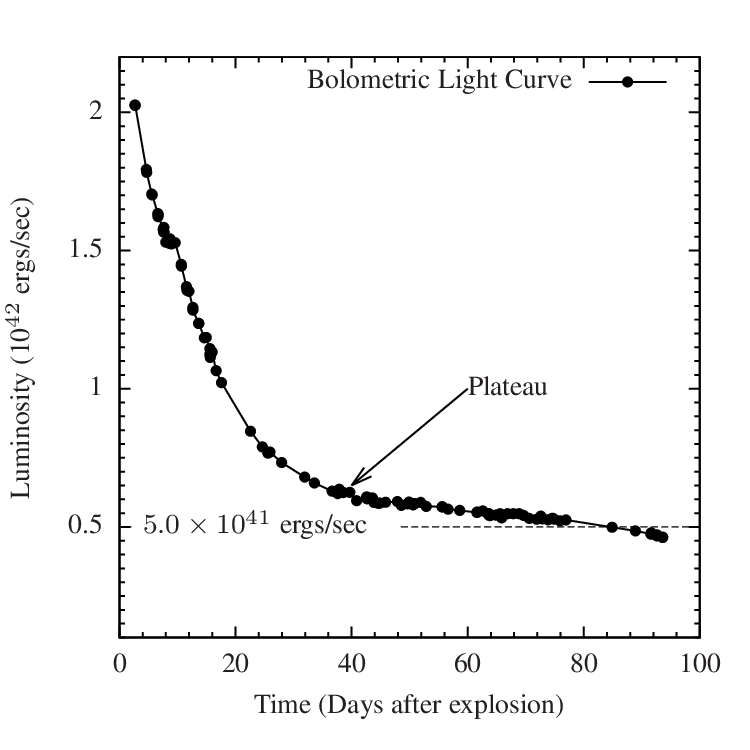}
\caption{Bolometric light curve of SN 2012aw calculated using the published UBVRI photometry
from \citet{2013ApJ...764L..13B} \& \citet{2013NewA...20...30M}. The details of this procedure are
described in section 4.2. The value of plateau phase luminosity is around $\sim 5.0\times 10^{41}\ 
\rm erg\ s^{-1}$.}
\label{fig:bol-lc}
 \end{center}
\end{figure}

To get the Compton cooling timescale we need the bolometric luminosity. We construct a bolometric 
lightcurve using published photometric (UBVRI) data from \citet{2013ApJ...764L..13B} (\emph{Swift 
photometry}) \& \citet{2013NewA...20...30M}. We take the available photometric data and fill in the 
gaps using linear interpolation. Note that we have not included the infrared photometry which is 
not available at these epochs, and due to this the bolometric luminosity may be a higher by at most 
$\sim0.30\ \rm dex$ at the plateau phase. The swift photometry has been converted to flux from 
count 
rate using count to flux conversion factors from \citet{2008MNRAS.383..627P}. We calculate the 
bolometric lightcurve by integrating over the resulting photometric data using a simple trapezoidal 
integration rule. The calculated bolometric light curve is shown in Figure \ref{fig:bol-lc}. Late 
part ($t>200 \ \rm days$) of the bolometric light curve used in calculation is taken from 
\citet{2013MNRAS.tmp.1583B} who have also calculated the photospheric radius and temperature
evolution of SN 2012aw. The radiation density at radiosphere can be calculated as from 
\begin{equation}
u_{rad}(t)=\dfrac{L_{Bol}(t)}{4\pi R(t)^2 c}
\end{equation}
where $R(t)$ is the radius of the radiosphere (forward shock) at a given time and is given as
\begin{equation}
 R(t) =R_0\left(\dfrac{t}{t_0}\right)^m {\label{eqn-radius}}
\end{equation}

\subsection{Inverse Compton Vs. Synchrotron Cooling}
\begin{figure}
 \begin{center}
   \includegraphics[width=0.8\textwidth]{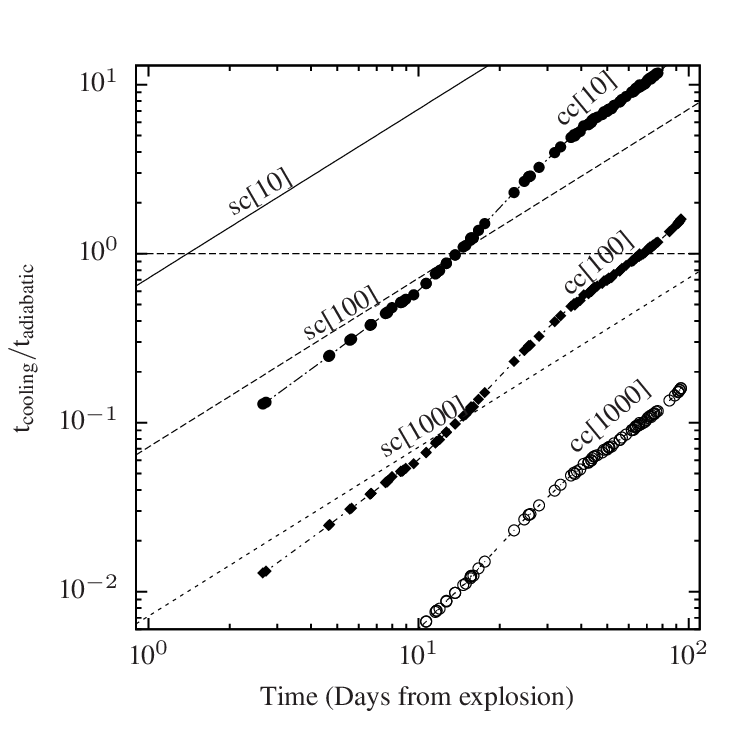}
\caption{Cooling timescales for electrons calculated using the computed bolometric light curve 
for various values of Lorentz factors as marked in the diagram by sc[$\gamma_i$] for synchrotron 
Cooling \& cc[$\gamma_i$] for Compton cooling respectively. Note that Compton cooling process is 
dominant over the  synchrotron cooling process at any given Lorentz factor $\gamma_i$. Also both 
inverse Compton cooling and synchrotron cooling dominate over the adiabatic cooling.}
\label{fig:cooling-ts1}
 \end{center}
\end{figure}

The calculated inverse Compton and synchrotron cooling timescales for electrons of different 
Lorentz factor $\gamma_i$ are shown in the Figure \ref{fig:cooling-ts1} in comparison to the 
adiabatic timescale. At all values of $\gamma_i$, inverse Compton cooling timescale is very small 
compared to synchrotron cooling timescale.

The ratio of synchrotron and Compton cooling timescales is independent of electron energy
\begin{equation}
 \dfrac{t_{SC}}{t_{IC}}\propto \dfrac{u_{rad}}{u_B}
\end{equation}

\begin{figure}
 \begin{center}
  \includegraphics[width=0.8\textwidth]{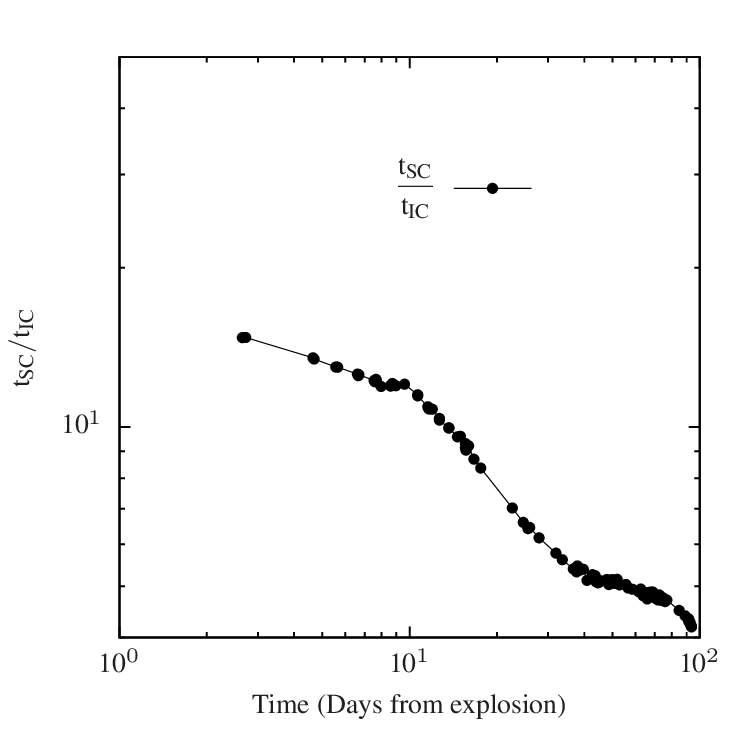}
  \caption{Ratio of synchrotron and Compton cooling timescales for electrons calculated using 
  the computed bolometric light curve. Note that in case of SN 2012aw Compton cooling process is 
  dominant over the  synchrotron cooling process because $t_{Comp}\ll t_{Sync}$. After around $100\ 
  \rm days$ the object enters the regime where synchrotron cooling is dominant over inverse Compton 
  cooling, but by that age adiabatic expansion losses are the most important (see Figure 
\ref{fig:cooling-ts1}).}
\label{fig:cooling-ts2}
 \end{center}
\end{figure}

It is evident from Figure \ref{fig:cooling-ts2} that Compton cooling dominates over the synchrotron 
cooling mechanism. Therefore in order to model the radio spectrum at early epochs and at high 
frequency we need to consider the effect of inverse Compton cooling mechanism on emission. This can 
be done by modeling the  kinetic equation for electrons with the relevant energy loss terms 
included.
\subsection{Cooling Frequency}
Assuming that an electron emits synchrotron radiation at its characteristic frequency $\nu_c$, we 
can get an estimate of frequencies which are affected at a given age by comparing the adiabatic 
timescale $t_{ad}\sim 1.5t$ and Compton cooling timescale $t_{IC}$. Electrons which are affected by 
cooling ($t_{Comp}<t_{ad}$) have energy greater than
\begin{equation}
 E>\dfrac{1}{3.97\times 10^{-2}u_{rad}\times1.5t}
\end{equation}
Using $\nu_c\sim c_1BE^2$, where $c_1$ is a constant, the minimum frequency above which 
effects due to Compton cooling are present can be written as

\begin{eqnarray}
 \nu_{min} &\gtrsim& c_1\times B_0\left(\frac{t_0}{t}\right)\left(\dfrac{4\pi R_0^2 c 
t}{5.96\times   10^{-2}t_0^2L_{bol}}\right)^2 \\
 \nu_{min}&=& \dfrac{c_1B_0}{t_0^3}\times \left(\dfrac{4\pi R_0^2c}{5.96\times 10^{-2}}
 \right)^2\times\left(\dfrac{t}{L_{bol}^2}\right)\\
 &=&0.78\left(\dfrac{t}{10\ \rm days}\right)\left(\dfrac{L_{Bol}}{10^{42}}\right)^{-2} \ \rm GHz 
 \label{eqn:c-cool-freq}
\end{eqnarray}

\begin{figure}
 \begin{center}
  \includegraphics[width=0.8\textwidth]{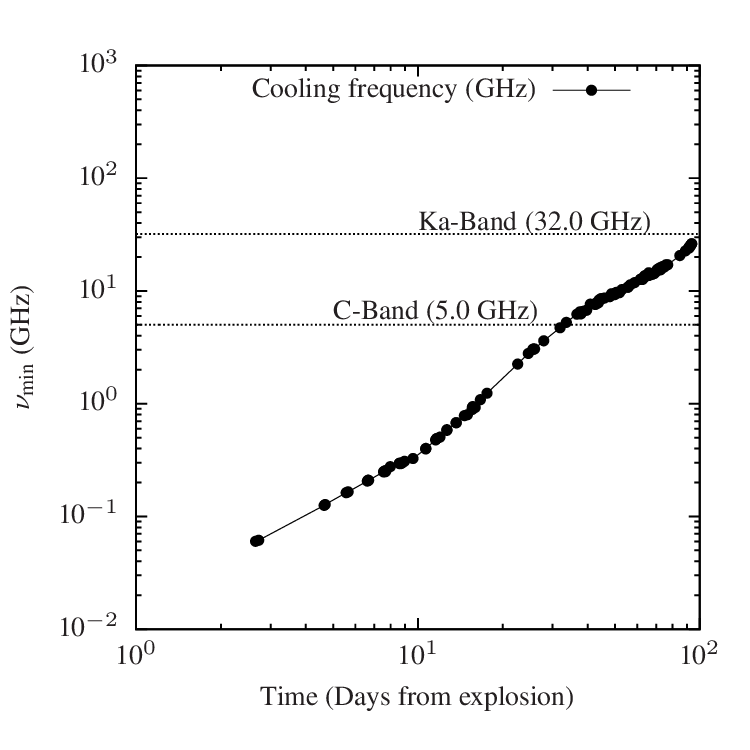}
  \caption{The minimum frequency affected by cooling plotted as a function of the age 
of SN 2012aw. Note that the dotted lines at $32.0\ \rm GHz$ and $5.0\ \rm GHz$ show that at early 
time all the VLA bands will have some effect of electron cooling. For the C-Band the cooling phase 
lasts till $\sim 30\ \rm days$. The rate of energy loss by relativistic electrons due to Compton 
scattering is proportional to the radiation density. During the plateau phase of a Type II-P 
supernova, the bolometric luminosity remains high for an extended period of time. This leads to 
electron cooling. The rate of cooling is proportional to square of electron energy, therefore the 
higher energy electrons cool much faster than the lower energy electrons. Therefore the higher 
frequency bands are affected relatively more compared to lower frequency bands.}
\label{fig:cool-freq}
 \end{center}
\end{figure}

The minimum frequency which is affected by cooling is shown in Figure \ref{fig:cool-freq}. It shows 
that at very early times most of the \emph{JVLA} radio bands are affected, but as the SN bolometric 
flux decreases $\nu_{min}$ goes to larger and larger values as can be seen from Equation 
\ref{eqn:c-cool-freq}. It shows that electron cooling needs to be considered for a self-consistent 
modeling of early times high frequency radio emission.

\section{ Cooling Affected Electron Population}
In order to evaluate the effect of electron cooling on radio emission, we can solve the full 
electron kinetic equation numerically and calculate the fluxes at any given time from the resulting 
electron distribution. The rate of change of energy of an electron is given by
\begin{equation}
 \dfrac{dE}{dt}=\left(\dfrac{dE}{dt}\right)_+-\left(\dfrac{dE}{dt}\right)_{-}
\end{equation}
where `$+$' and `$-$' represent energy gain and loss processes. At an energy $E_{max}$, both rates 
can become equal and the electron can not be accelerated further. We therefore obtain an electron 
distribution which is bounded at the higher energy end. The cutoff is dependent on the bolometric 
luminosity and the radius of the forward shock. We can get the upper limit on energy by comparing 
the cooling timescale $t_{Comp}$ and average acceleration timescale $\tilde t_{acc}$ (it quantifies 
the time required for electron to be accelerated to a given energy) for radio emitting electrons.
The condition for $E_{max}$ is
\begin{equation}
 \dfrac{t_{Comp}}{\tilde t_{acc}}< 1
\end{equation}
The inequality gives $E_{max}$ in terms of bolometric luminosity, forward shock radius and 
$\tilde t_{acc}$ as a function of time
\begin{equation}
 E_{max}=\dfrac{4 \pi R(t)^2 c}{3.97\times 10^{-2}\tilde t_{acc}L_{bol}(t)}
\end{equation}
We truncate the original power law electron distribution at $E_{max}$. The electron distribution at 
a time $t$ can be written \citep{1970ranp.book.....P} as
\begin{equation}
N(E,t)=\left\{
\begin{array}{ll}
N_0E^{-\gamma}\left(1-\dfrac{E}{E_{max}}\right)^{\gamma-2} \ &E_{min}< E <E_{max} \\\\
0 \quad & E>E_{max}
\end{array}
\right.
\label{eqn:elecdist}
\end{equation}
where
\begin{equation}
E_{min}=m_ec^2
\end{equation}
and $N_0$ is the normalization of the original distribution \citep{1998ApJ...499..810C} and is 
related to the equipartition factor.
\begin{equation}
 N_0=\dfrac{\tilde \alpha (\gamma-2)B^2E_{min}^{\gamma-2}}{8\pi}
\end{equation}

\section{Calculating Radio Spectrum}
To obtain the emission coefficient ($\epsilon_\nu$) and absorption
coefficient ($\kappa_\nu$) using the modified electron
population we use the equations for $\epsilon_\nu$ and $\kappa_\nu$ from Pacholczyk (1970) as
\begin{eqnarray}
 \epsilon_\nu&=&c_3 H \sin \theta \int_{E_{min}}^\infty N(E)F(x)\ dE \label{eqn:emiss} \\
 \kappa_\nu&=&-\dfrac{c^2}{2\nu^2}c_3 H\sin \theta\int_{E_{min}}^\infty E^2 \dfrac{d}{dE}
 \left(\dfrac{N(E)}{E^2}\right)F(x)\ dE \label{eqn:absor}
\end{eqnarray}
where
\begin{eqnarray}
 x&=&\dfrac{\nu}{\nu_c}\\
 \nu_c&=&c_1 H\sin \theta E^2\\
 F(x)&=&x\int_x^\infty K_{5/3}(z)\ dz
\end{eqnarray}
We can get the emission and absorption coefficient by substituting
Equation \ref{eqn:elecdist} into Equation \ref{eqn:emiss} \& \ref{eqn:absor}. We can write $E$
as a function of $x$ using $\nu_c=c_1 H\sin \theta E^2$ as
\begin{eqnarray}
 E&=&\dfrac{A}{\sqrt{x}}\\
 dE&=&-\dfrac{1}{2}x^{-3/2}A\ dx\\
 \text{where,}\ A&\equiv&\left(\dfrac{\nu}{c_1H\sin \theta}\right)^{1/2}
\end{eqnarray}
The equations for emission and absorption coefficient after substitution become
\begin{eqnarray}
 \epsilon_\nu&=&-\dfrac{1}{2}c_3N_0\left(\dfrac{\nu}{c_1}\right)A^{-(\gamma+1)}I_0\\
 \kappa_\nu&=&-\dfrac{c^2}{4\nu^2}c_3N_0\left(\dfrac{\nu}{c_1}\right)A^{-\gamma}\left(A^{-2}I_1+
 A^{-1}I_2\right)
\end{eqnarray}
The integral in the above formula are as following
\begin{eqnarray}
 I_0&=&\int_{x_1}^{x_2} x^{(\gamma-3)/2}g(x)^{\gamma-2}F(x)\ dx\\
 I_1&=& (\gamma+2)\int_{x_1}^{x_2} x^{(\gamma-2)/2}g(x)^{\gamma-2}F(x) dx\\
 I_2&=& \dfrac{(\gamma-2)}{E_{max}}\int_{x_1}^{x_2} x^{(\gamma-3)/2}g(x)^{\gamma-3}F(x) dx\\
 &&\text{where,}\ g(x)=\left(1-\dfrac{Ax^{-1/2}}{E_{max}}\right)
\end{eqnarray}
The limits of integration are given by
\begin{eqnarray}
x_1&=&\left(\dfrac{A}{E_{min}}\right)^2\\
x_2&=&\left(\dfrac{A}{E_{max}}\right)^2
\end{eqnarray}
The source function is defined as
\begin{equation}
 S_\nu=\dfrac{\epsilon_\nu}{\kappa_\nu}
\end{equation}
For our case it becomes
\begin{equation}
S_\nu=\dfrac{2\nu^2}{c^2} \dfrac{AI_0}{\left(I_1+AI_2\right)}
\end{equation}
The radiative transfer problem can be easily solved for the case of a planar emission region of 
thickness, $s$
\begin{equation}
 \pi R^2s=f\dfrac{4\pi}{3}R^3
\end{equation}
where $f$ is the filling fraction. We use $f=0.5$ in our calculation. The radiative transfer 
equation is
\begin{equation}
 \dfrac{dI_\nu}{d\tau_\nu}=I_\nu-S_\nu
\end{equation}
It can be integrated simply in case of a homogeneous emission region from $0$ to $s$ as 
\begin{equation}
 I_\nu(s)=I_\nu(0)e^{-\tau_\nu(s,0)}+\int_0^{s} \kappa_\nu S_\nu  e^{-\tau_\nu(s,s')}\ ds'
\end{equation}
As there is no incident radiation at $s=0$, therefore $I_\nu(0)=0$ and the solution becomes
\begin{equation}
 I_\nu(\tau_\nu)=S_\nu(1-e^{-\tau_\nu})
\end{equation}
where $\tau_\nu$ is defined as the optical depth as following
\begin{equation}
 \tau_\nu=\int_0^s \kappa_\nu\ ds=s\kappa_\nu
\end{equation}
The flux can be calculated by integrating $I_\nu$ over the solid angle $\Omega$ as
\begin{equation}
 F_{\nu}=\int I_{\nu}d\Omega=S_\nu(1-e^{-\tau_\nu})\Omega
\end{equation}
The integral for emission and absorption coefficient are evaluated numerically to
obtain the radio light curves. The effect of FFA \citep{1998ApJ...499..810C} can be included as
\begin{equation}
 F_{\nu}=S_{\nu}\Omega (1-e^{-\tau_\nu})\times\exp\left\{-\left(\dfrac{t}{t_{ff}}\right)^{-3}
 \left(\dfrac{\nu}{\nu_1}\right)^{-2.1}\right\}
\end{equation}
where $t_{ff}$ is the time at which the optical depth to FFA becomes unity at frequency $\nu_1$. In 
the calculation we have used $\nu_1= 3 \ \rm GHz$. 

\section{Results of Modeling the Radio Observations}
\begin{figure}
\begin{center}
 \includegraphics[width=0.8\textwidth]{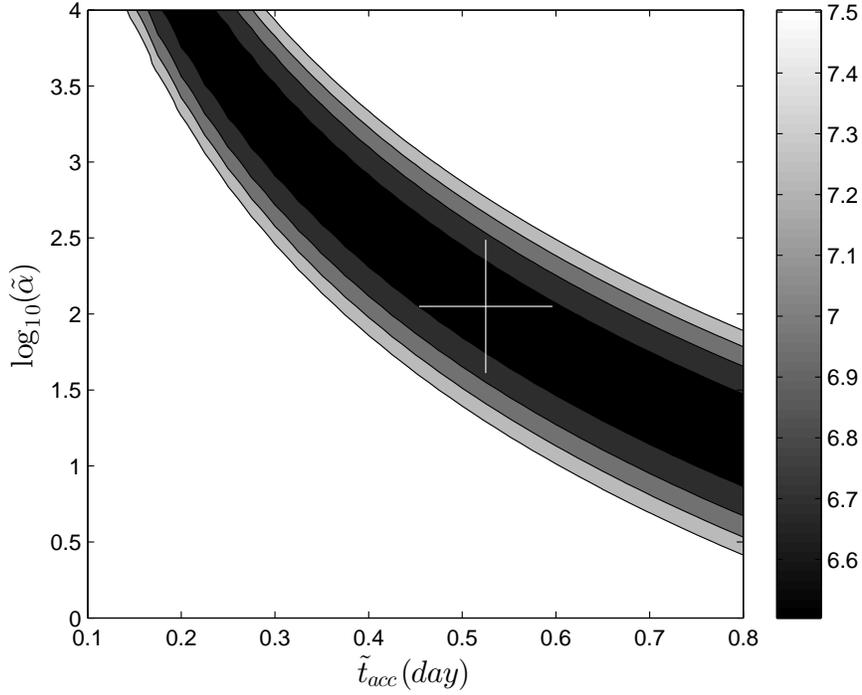}
 \caption{A plot of calculated $\chi^2_{\nu}$ as a function of $\tilde t_{acc}$ and 
equipartition
  factor ($\log_{10}\tilde\alpha$) for $t_{ff}=18.5 \ \rm days$ calculated using Model-4 (refer 
  to Table 2) with the explosion date 2012 Mar 15.3 UT. The cross represents the best fit value of 
  $\tilde t_{acc}$ and $\tilde \alpha$ with $\chi^2_\nu\sim6.50$. The gray scale shows the value of 
  $\chi_{\nu}^2$ and the corresponding contours. The levels marked in contour plot are separated by 
  $\sim 0.2$. For Compton cooling  to be dominant we need $\tilde \alpha>1.0$, therefore the region 
  below $\log_{10}(\tilde \alpha)<0.0$ is rejected.  The best fit parameters of this model are 
  given in Table \ref{Table-2: modeling}.} \label{fig:contour_plot}
 \end{center}
\end{figure}

Using the model described above we compute the radio fluxes and fit them to the 
observations as follows:
\begin{enumerate}
 \item For a given explosion date ($t_{ex}$), fit the $3\ \rm GHz$ radio light curve with an SSA 
 model to obtain $m$, $F_{p}$ and $t_{p}$.
 \item Calculate the radius ($R_p$) and magnetic field ($B_p$) estimates  \item Use the $R_p$, 
$B_p$ 
 and $t_p$ in the cooling model (Model-3 \& -4) to obtain the best fit values of $t_{ff}$, 
$t_{acc}$ 
 and $\log_{10}(\tilde \alpha)$ based on $\chi^2$ minimization.
 \item In computing model-3 \&4, we use the optical light curve properly referenced according to the
 explosion date.
\item Compare models for different value of explosion date.
\end{enumerate}
Using the above procedure, we calculate best fit parameters by minimizing $\chi^2_{\nu}$ over the 
3-dimensional parameter space using Model-4 (Table 2). We use the {\it S} ($3.0\ \rm GHz$), {\it C} 
($5.0\ \rm  GHz$), {\it X} ($8.5\ \rm  GHz$) \& {\it K} ($21.0\ \rm  GHz$) band data for fitting 
purposes. The {\it Ka} ($32.0\ \rm  GHz$) band observations are consistent with its light curve 
computed from the parameters obtained from fitting the other frequencies. The resolution of the 
grid is $0.05$ on $\log_{10}(\tilde \alpha)$ axis, $0.5 \ \rm days$ on the $t_{ff}$ axis and $0.025 
\ \rm days$ on the $\tilde t_{acc}$ axis. For Compton cooling to be dominant we need $\tilde 
\alpha>1.0$, therefore  the region below $\log_{10}(\tilde \alpha)<0.0$ is rejected. We obtain 
the best fit value of $t_{ff}=18.5\ \rm days$, $\tilde t_{acc} = 0.53 \ \rm days$ and $\tilde \alpha 
= 1.12\times 10^2$ for the parameters. The $\chi_\nu^2$ corresponding to these parameter values is 
$6.5$. The contour plot visualizing the $\log_{10}\tilde \alpha - \tilde t_{acc}$ space is shown in 
Figure \ref{fig:contour_plot}. The levels marked in contour plot are separated by $\sim 0.2$. 
Because of the weak dependence of observed quantities on $\tilde \alpha$, it is not very strongly 
constrained by the radio observations alone. The values of fitted parameters are reported in Table 
2.

Another estimate of $\tilde \alpha$ can be obtained by using the observed X-ray 
luminosity as the upper limit of the IC contribution to X-ray luminosity\footnote{The X-ray 
luminosity equations assume that the circumstellar medium is formed by winds with constant 
parameters.
\begin{equation}
L^{X}_{obs}=L^{X}_{IC}+L^{X}_{Thermal} \Rightarrow L^{X}_{obs}\gtrsim L^{X}_{IC}
\end{equation}
Using the expression for $E(dL^{X}_{IC}/dE)$ from \citet{2012ApJ...761..100C}
and integrating it over the energy range $0.2\ \rm Kev$ to $2.0\ \rm Kev$
, we get
\begin{equation}
 8.8\times10^{36}\gamma_{min}S_{\star}\tilde \alpha^{11/19}V_{s4}\left(\dfrac{L_{bol}(t)}{10^{42}\ 
 \rm erg s^{-1}}\right)\left(\dfrac{t}
 {10 \ \rm days}\right)^{-1}\lesssim L^{X}_{obs} \label{eqn:IC-flux}
\end{equation}
where $\gamma_{min}$ is the minimum Lorentz factor of electrons and $S_\star$ is the radio emission 
measure given by Equation 14 of \citet{2012ApJ...761..100C}
\begin{equation}
 S_{\star}=1.0\left(\dfrac{f}{0.5}\right)^{-8/19}\left(\dfrac{F_{\nu p}}{\rm mJy}\right)^{-4/19}
 \left(\dfrac{D}{\rm Mpc}\right)^{-8/19}\left(\dfrac{\nu}{5\ \rm GHz}\right)^{2}\left(\dfrac{t}{10\ 
 \rm days}\right)^2
\end{equation}
Using $f=0.5$, we get $S_{\star}=3.97$ for SN 2012aw. The value of $L^{X}_{obs}$ at an age of $\sim 
5.6\ \rm days$ is taken from \citet{2012ATel.3995....1I}. Substituting $S_{\star}$ and $V_{s4}\sim 
R0/t0$ and $L_{bol}=1.7\times 10^{42}$
into Equation \ref{eqn:IC-flux} gives
\begin{equation}
\tilde \alpha \sim 0.35_{-0.24}^{+1.14}\times 10^2
\end{equation}
This is smaller than the value of $\tilde \alpha$ giving the best fit to the radio data but is 
consistent with the later within error limits (refer Table \ref{Table-2: modeling}).
\emph{Chandra} observed the field of SN 2012aw on 2012 Apr 11. We analyzed the data and determine 
an X-ray luminosity $(6.0 \pm 1.4)\times 10^{37} \ \rm  erg/sec/keV$ at $1.0 \ \rm keV$. This 
implies an $\tilde \alpha \sim 0.22_{-0.13}^{+0.45}\times 10^2$.} 

\begin{figure}
\begin{center}
  \includegraphics[width=0.8\textwidth]{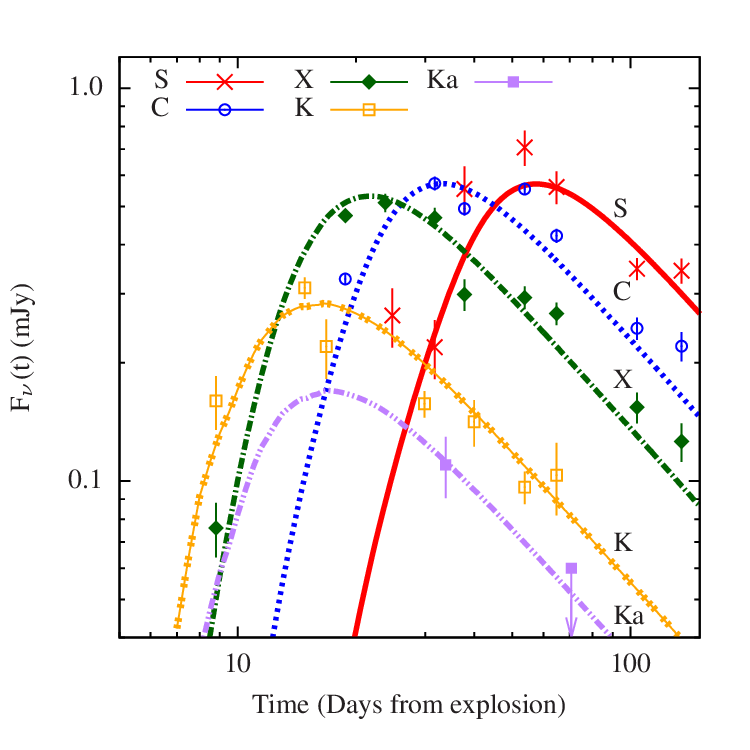}
  \caption{Radio observations of SN 2012aw and computed light curves based on Model-4 (Table 2).
  The parameters of the plotted model are: $t_{ff}=18.5\ \rm days$, $\tilde t_{acc}=0.53\ \rm days$ 
  and  equipartition factor $\tilde \alpha\sim 1.12\times 10^2$. The bands are as follows: 
{\it S}   ($3.0\ \rm GHz$), {\it C} ($5.0\ \rm  GHz$), {\it X} ($8.5\ \rm  GHz$), {\it K} ($21.0\ 
\rm    GHz$) \& {\it Ka} ($32.0\ \rm  GHz$).}\label{fig:fit-rlc}
 \end{center}
\end{figure}

We note that the magnetic field and relativistic electrons are away from equipartition 
regime. The value of $t_{ff}$ can be used to get the $\dot M/v_w$ by inverting Equation 4 of 
\citet{2006ApJ...641.1029C}
\begin{equation}
\dfrac{\dot M_{-6}}{v_{w1}} \approx \left(\dfrac{t_{ff}T_{cs5}^{1/2}V_{s4}}{6}\right)^{3/2}
\left(\dfrac{\nu}{8.46\ \rm GHz}\right)
\end{equation}
Using $V_{s}\sim R_0/t_0$, $T_{cs5}\sim 1.0$ and $t_{ff}=18.5\ \rm days$ at $\nu_1=3\ \rm GHz$, 
we get
\begin{equation}
\dot M_{-6}/v_{w1}\sim1.9
\end{equation}
The calculated radio light curves for the best fit parameters are shown in Figure \ref{fig:fit-rlc}.
Using our model we are able to explain the early time data at high frequency.

We also model the effect of varying the explosion date, $t_{ex}$ since there is a time difference 
of $1.6\ \rm days$ between the last non-detection \citep{2012ATel.3996....1P} and the first optical 
detection of the SN \citep{2012CBET.3054....1F}. The explosion date affects the calculation of 
radio flux especially at high frequencies, since the relativistic electrons experience different 
radiation environments due to the change of the density of \emph{UVOIR} photons at the radiosphere. 
We calculate radio fluxes due to synchrotron emission by electrons for different explosion dates 
and fit the fluxes to the observed radio data. The results are summarized in Table \ref{Table-3: 
explosion-date-model} for Model-4 (refer to Table 2). The best fit is obtained for $t_{ex}$ of 2012 
March 15.3.

\section{Conclusions \& Discussion}
We have reported the radio observations of SN 2012aw which has already been studied well in the 
optical and UV bands. Our observations spanning $184 \ \rm days$ make it one of the best observed 
Type II-P radio  SN. We find that the spectral index values are smaller than the values 
expected for optically thin regime. We interpret this as a signature of electron cooling at a young 
age. Specifically we find that inverse Compton cooling dominates over the synchrotron cooling 
process in the case of SN 2012aw. Although \citet{2006ApJ...641.1029C} had predicted the effect of 
electron cooling on radio light curves, this is the first unambiguous evidence of cooling of 
relativistic electrons in a young supernova due to inverse Compton scattering of low energy photons. 
We consider the effects of Compton cooling in order to  self-consistently model the high frequency 
radio emission. We fit the radio data  to the model and estimate its parameters. We find that 
radiating plasma is away from equipartition ($\tilde \alpha \sim 1.12\times 10^{2}$) and 
relativistic electrons carry a greater fraction of the thermal energy compared to the post-shock 
magnetic field. A similar result has been noted in case of SN 2011dh (a Type IIb SN)
by \citet{2013MNRAS.tmp.2312H} for which $\tilde \alpha \sim 10^3$, which implies $\epsilon_e\gg 
\epsilon_B$,  energy density in relativistic electrons exceeds the energy density in magnetic field 
(\citet{2012ApJ...752...78S} have noted a value of $\tilde \alpha \sim 30$ for the case of SN 
2011dh). The case of SN 1993J (another Type IIb) is in contrast to SN 2011dh as in the former case 
\citet{2004ApJ...612..974C} \& \citet{1998ApJ...509..861F} noted that the equiparition factor is 
$\sim 10 ^{-4}$ ($\epsilon_e\ll \epsilon_B$, energy density in magnetic field exceeds the energy 
density in relativistic electrons). We determine the value of $\dot M\sim 1.9\times10^{-6}\ \rm 
M_\odot\ yr ^{-1}$ and it is consistent with the empirically estimated mass loss rate for red giant 
progenitors of Type II-P SNe \citep{1977A&A....61..217R,1988A&AS...72..259D}. To investigate 
phenomenon associated with electron cooling, observations of radio bright  SN at young age in high 
frequency bands using \emph{ALMA} and/or \emph{CARMA} will be needed as has been done 
in the case of SN 2011dh by \citet{2013MNRAS.tmp.2312H} and \citet{2012ApJ...752...78S}. Good 
quality early X-ray observations by \emph{Swift} and/or \emph{Chandra} are crucial to get stringent 
limits on equipartition factor (including independent estimates on $\epsilon_e$ and $\epsilon_B$) 
and the contribution of thermal emission to the X-ray flux, as has been done in the case of SN 
2004dj by \citet{2012ApJ...761..100C}.

\section{Acknowledgments}
Initial observation of SN 2012aw was requested under Joint Chandra-NRAO Cycle-13 ToO proposal
(Proposal ID No. 13500809) on Type II-P SNe. AR wishes to thank the Department of Physics at the 
West Virginia University for hospitality during proposal development. We thank the anonymous 
reviewer for his/her comments, queries and suggestions, which significantly helped us in 
improving the manuscript. We wish to acknowledge the support of TIFR 12th Five Year Plan (Project 
No: 12P-0261). We would like to thank the Director \emph{Karl G. Jansky VLA} for granting us the 
observations under Director's Discretionary Time. NY wishes to acknowledge the support of CSIR-SPM 
fellowship (SPM-07/858(0057)/2009-EMR-I).

\begin{deluxetable}{lcccc}
\tablecolumns{5}
\tablewidth{0pt}
\tablenum{1}
\tablecaption{Radio Observations of SN 2012aw. The explosion date is taken to be 2012 Mar 16.1(UT).}
\label{Table-1: Radio Obs}
\tablehead{ \colhead{$Date(UT)$} & \colhead{$\Delta t\ \rm(days)$} & \colhead{$\nu\ \rm(GHz)$} &
\colhead{$f_{\nu}\ \rm(\mu Jy)$} & \colhead{$r.m.s \ \rm(\mu Jy)$}}
\startdata
Mar24.10 & 8.00 & 8.5 &  76 & 12.0\\
Mar24.10 & 8.00 & 20.8 & 160 & 25.0\\
Mar30.09 & 14.00 & 21.2 & 310 & 19.6\\
Apr01.31 & 16.00 & 20.8 & 220 & 38.0\\
Apr03.08 & 17.98 & 8.5 & 474 & 11.0\\
Apr03.08 & 17.98 & 5.0 & 327 & 11.0\\
Apr08.04 & 22.94 & 8.9 & 510 & 27.4\\
Apr09.04 & 23.94 & 5.5 & 559 & 42.7\\
Apr09.04 & 23.94 & 2.9 & 264 & 45.4\\
Apr09.04 & 23.94 & 1.5 & $<$150.0 & 50.0\\
Apr14.19 & 29.09 & 21.2 & 157 & 12.2\\
Apr16.02 & 30.92 & 3.0 & 219 & 37.6\\
Apr16.02 & 30.92 & 5.5 & 572 & 23.1\\
Apr16.02 & 30.92 & 8.9 & 468 & 27.8\\
Apr18.12 & 33.02 & 32.0 & 110 & 19.5\\
Apr22.05 & 36.95 & 3.1 & 554 & 78.4\\
Apr22.05 & 36.95 & 5.5 & 493 & 19.2\\
Apr22.05 & 36.95 & 9.0 & 299 & 27.6\\
Apr24.27 & 39.17 & 21.2 & 142 & 19.1\\
May08.06 & 52.94 & 3.0 & 707 & 72.8\\
May08.06 & 52.94 & 5.5 & 554 & 17.6\\
May08.06 & 52.94 & 9.0 & 293 & 19.5\\
May08.06 & 52.94 & 21.2 & 96 & 9.2\\
May19.10 & 64.00 & 3.2 & 560 & 53.6\\
May19.10 & 64.00 & 5.5 & 421 & 17.3\\
May19.10 & 64.00 & 9.0 & 267 & 17.7\\
May19.10 & 64.00 & 21.2 & 103 & 21.6\\
May23.19 & 68.09 & 32.0 & $<$60.0($4\sigma$) &15.0\\
Jun27.10 & 103.00 & 3.2 & 347 & 22.2\\
Jun27.10 & 103.00 & 5.5 & 245 & 15.9\\
Jun27.10 & 103.00 & 9.0 & 154 & 13.9\\
Jul28.03 & 133.97 & 3.2 & 343 & 25.0\\
Jul28.03 & 133.97 & 5.5 & 220 & 18.9\\
Jul28.03 & 133.97 & 8.9 & 126 & 14.2\\
Sep15.00$^\dag$ & 182.90 & 1.3 & 436 & 81.0\\
\enddata
\tablenotetext{\dag}{GMRT data point}
\end{deluxetable}

\begin{deluxetable}{cccc}
\tablecolumns{4}
\tablenum{2}
\tablewidth{0pt}
\tablecaption{Radio emission modeling results of SN 2012aw.}\label{Table-2: modeling}
\tablehead{\colhead{Model} &\colhead{Properties} &\colhead{Parameters} & \colhead{$\chi_{\nu}^2$}}
\startdata
 Model-1  &SSA &$m,t_{ff},f_p,t_p,\gamma$& 10.05  \\
 & No Cooling&$m=1.1^{\S}$&  \\
 & &$f_p(5\ \rm GHz)=0.52 \ mJy$& \\
 & &$t_p(5\ \rm GHz)=29.28\ day$& \\
 & &$\gamma=3.2$& \\
 & & {\bf Inconsistent with $m\lesssim1$} & \\
 Model-2  &SSA+FFA &$m,f_p,t_p,\gamma,t_{ff}$& 7.43 \\
 & No Cooling &$m=1.1^{\S}$& \\
 & &$f_p(5\ \rm GHz)=0.61 \ \rm mJy$& \\
 & &$t_p(5\ \rm GHz)=23.64\ \rm day$& \\
 & &$\gamma=3.1$& \\
 & &$t_{ff}(5\ \rm GHz)=15.84\ \rm day$& \\
 & & {\bf Inconsistent with $m\lesssim1$} & \\
 Model-3$^{\dag}$ &SSA+Cooling &$\tilde t_{acc},\tilde \alpha$ & 7.45\\
  & &$m=0.97^{\ddag}$& \\
  & &$\gamma=3.1 \ \rm (fixed)$& \\
  & &$\tilde t_{acc}=0.55^{+0.20}_{-0.15}\ \rm day$& \\
  & &$\log_{10}\tilde \alpha=2.70^{+0.65}_{-0.70}$& \\
 Model-4$^{\dag}$ &SSA+FFA+Cooling &$t_{ff},\tilde t_{acc},\tilde \alpha$  & 6.50 \\
  & &$m=0.97^{ \ddag}$& \\
  & &$\gamma=3.1 \ \rm (fixed)$& \\
  & &$t_{ff}(3\ \rm GHz)=18.5^{+0.5}_{-0.5}\ \rm day$& \\
  & &$\tilde t_{acc}=0.53^{+0.23}_{-0.18}\ \rm day$& \\
  & &$\log_{10}\tilde \alpha=2.04^{+0.8}_{-0.6}$& \\
  \enddata
\tablenotetext{\S}{Based on fit to the multi-band radio data and explosion date 2012 Mar 16.10 UT.}
\tablenotetext{\dag}{For these models we use the radius and magnetic field determined using the
$3\ \rm GHz$ lightcurve. Refer to Equations \ref{eqn-magnetic_field} and \ref{eqn-radius}.}
\tablenotetext{\ddag}{Based on fitting the $3\ \rm GHz$ radio light curve.}
\end{deluxetable}

\begin{deluxetable}{cccccc}
\tablecolumns{6}
\tablenum{3}
\tablewidth{0pt}
\tablecaption{Model fits with respect to the fiducial date of explosion for Model-4$^\dag$} 
\label{Table-3: explosion-date-model}
\tablehead{\colhead{Explosion Date} &\colhead{$\chi_\nu^2$} &\colhead{$m^\S$}
&\colhead{$t_{ff}\ \rm (day)$} & \colhead{$\tilde t_{acc}\ \rm (day)$} & \colhead{$\log_{10}\tilde 
\alpha$}}
\startdata
 2012 Mar 15.30 & 6.50 & 0.97 & 18.5 & 0.53 & 2.04\\
 2012 Mar 16.10 & 6.76 & 0.98 & 17.0 & 0.90 & 0.69\\
 2012 Mar 16.90 & 7.16 & 0.99 & 15.0 & 0.70 & 1.00\\
\enddata
\tablenotetext{\S}{Based on fitting the $3\ \rm GHz$ radio light curve.}
\tablenotetext{\dag}{The earliest detection of the SN \citep{2012CBET.3054....1F} was on 2012 
Mar 16.9 UT, whereas the last reported non detection \citep{2012ATel.3996....1P} was on 2012 Mar 
15.3 UT.}
\end{deluxetable}

\bibliographystyle{apj}

\end{document}